\documentstyle[12pt,epsf]{article}

\renewcommand{\bar}[1]{\overline{#1}}
\newcommand{\etal}{{\em et al.}}
\newcommand{\ie}{{\it i.e.}}

\begin{document}

\begin{flushright}
SLAC--PUB--7445\\
April 1997
\end{flushright}

\thispagestyle{empty}
\flushbottom

\centerline{\Large\bf Looking for Exotic Multiquark States} \vspace{5pt}
\centerline{{\Large\bf in Nonleptonic B Decays}
\footnote{\baselineskip=14pt
Work supported in part by the Department of Energy under contract number
DE--AC03--76SF00515 and FAPESP and CNPQ.}}
\vspace{10pt}
\centerline{\bf S. J. Brodsky}
\vspace{5pt}
\centerline{\it Stanford Linear Accelerator Center}
\centerline{\it Stanford University, Stanford, California 94309}
\centerline{e-mail: sjbth@slac.stanford.edu} \vspace{5pt}
\centerline{and}
\vspace{5pt}
\centerline{\bf F. S. Navarra}
\vspace{5pt}
\centerline{\it Instituto de F\'{\i}sica, Universidade de S\~{a}o Paulo}
\centerline{\it C.P. 66318, 05315-970 S\~{a}o Paulo, SP, Brazil}
\centerline{e-mail: navarra@if.usp.br}
\vspace*{10pt}

{\baselineskip=16pt
\begin{abstract} Recent data on inclusive $B$ decays into charmonium
states taken by the CLEO collaboration suggest that a sizeable fraction
of the nonleptonic decays may consist of the $J/\psi\, \Lambda
\,\overline p$ three body final state, corresponding to a distinct
enhancement in the inclusive $J/\psi$ momentum distribution. The
kinematical boundary of this structure corresponds to the case where
the $J/\psi$ recoils nearly monoenergetically in the $B$ rest system
against a partner having mass of $\simeq 2$~ GeV. This may allow the
observation of a $\Lambda-\overline p $ bound state near or just below
threshold; \ie, strange baryonium, even if the production rate is
small. Using a phase space approach to the $B$ meson decay, we study
the $J/\psi$ momentum distribution and the effect of baryonium
formation. We also discuss the possible observation of pentaquarks and
hadronically-bound $J/\psi$ by the observation of monoenergetic baryons
in these decays.
\end{abstract}}


\newpage

\section{Introduction}

One of the most important new elements that quantum chromodynamics
brings to the strong interactions is the potential existence of new
bound states and resonances beyond the conventional mesons, baryons and
nuclei. These include exotic states such as gluonium ($gg$, $ggg$),
hybrids ($ q \overline q g $) and dibaryons such as the H (udsuds), etc.

A particularly sensitive measure of QCD binding is nuclear-bound
quarkonium, \ie, a ($c - \overline c$) or ($ b - \overline b$) pair
bound to a nucleus. In contrast to the usual strongly interacting
few-body systems,  the hadrons in the ($c \bar c$) ($uud$)
meson-nucleon system have no valence quarks in common. However
multiple-gluon exchange (the QCD van der Waals force) may lead to the
formation of bound states \cite{bst}. Using the fact that the
characteristic size of the $J/\psi$ is much smaller than the typical
hadronic scale ($\Lambda_{QCD}^{-1}$), a QCD calculation based on a
multipole expansion was performed in \cite{mano}, leading to
nuclear-bound quarkonium binding energies of order $\simeq 8-11$ MeV.
In addition, attractive interactions may also arise from the
$J/\psi\,p\,\rightarrow\,D\, \Lambda_c $ interaction \cite{Miller}.
Long-range exchange interactions may arise in second order perturbation
theory if there are significant nonvalence ($Q \overline Q q \overline
q$) Fock states in the quarkonium wavefunction or intrinsic charm ($u
u d c \overline c$) Fock states in the nucleon. There is thus a
possibility of a strong $J/\psi$-nucleon resonance or bound state.

From the experimental point of view, there is evidence for the state
$B_{\phi}= |q\,q\,q\,s\,\overline s >$ (where $q$ are $u$ or $d$
quarks), reported by the SPHINX collaboration \cite{vav},  and there
are proposals for a search for the $B_{\psi} = |q\,q\,q\,c\,\overline c
>$ with the same setup \cite{land} and at Fermilab as well \cite{moin}.

Another interesting type of exotic state that could arise in QCD are
bound states of baryons and antibaryons (baryonia) such as $ p -
\overline p$ and $\Lambda - \overline p,$ and pentaquarks, $P_Q$,
baryons containing five quarks like $u\,u\,d\,\overline c\,s$ or
$u\,u\,d\,c\,\overline s$ \cite{moin,lip}. Baryonia may have binding
energies as small as a few MeV as is characteristic of nuclear
interactions, but they are unstable and decay into mesons via $ q -
\overline q$ annihilation with possibly narrow widths. Such states have
been subject of intensive search\cite{ven}. In particular, experiments
have looked for strange baryonium in hadronic collisions, typically in
$K\,\overline p \rightarrow \Lambda\,\overline p\,p$ reactions at
laboratory energies of $ 8.25 $~GeV \cite{bau}, $ 18.5 $~GeV \cite{arm}
and $ 50 $~GeV \cite{cle}. The results concerning the width of these
states have been somewhat controversial. Some experimental searches
\cite{bau,apo} found $\Gamma \simeq 20 $ MeV whereas others
\cite{cle,arm} found $\Gamma \simeq 150 $ MeV. The existence of narrow
structures has never been established or completely ruled out.

Searches for the pentaquarks are in progress at Fermilab. Data from the
collaboration E-791 are currently being analyzed, and another
experiment, the E-781, is accumulating events with charmed baryons,
which may determine the existence of certain pentaquarks \cite{esc}.

In this paper we will show how to look for QCD exotic states in a
B-factory. In particular, we will show how the $B^-\rightarrow J/\psi\,
\Lambda \,\overline p $ decay channel can be used to search for the ($
J/\psi-\overline p $), ($ J/\psi-\Lambda $) and ($ \Lambda-\overline p
$) bound states. Also the appearance of two-body structures in the
momentum distribution of the $\overline \Lambda_c$ near $p = 0.47$
GeV/c in the $B^+\rightarrow \overline \Lambda_c + X$ decay channel may
provide a signature for the $u\,u\,d\,c\,\overline s$ pentaquark.

\section{$ B^-\rightarrow J/\psi\, \Lambda \,\overline p $ decay mode}

Among the numerous decay modes of the $B$ mesons, the $B \rightarrow
{\rm charmonium} + X$ channels have drawn particular attention. Recent
measurements of these decays performed by the CLEO collaboration
\cite{cleo} have found the following inclusive branching fractions:
$(1.12\pm 0.04 \pm 0.06) \%$ for $B\rightarrow J/\psi X$, $(0.34\pm
0.04 \pm 0.03) \%$ for $B\rightarrow \psi^{'} X$ and $(0.40\pm 0.06 \pm
0.04) \%$ for $B\rightarrow \chi_{c1} X$. Crossing this information
with the measured \cite{pdb} branching fraction $(2.5 \pm 0.4)\%$ for
$B\rightarrow \Lambda\,\overline p X $, one can infer that a
significant fraction of the charmonium decays correspond to the
three-body decay:
\begin{eqnarray}
B^-\rightarrow J/\psi\, \Lambda \,\overline p
\label{eq:bran}
\end{eqnarray}
This process, ideal for our purposes, is depicted in Fig. 1a. In first
place, the $ e^-e^+\rightarrow B \overline B$ environment is much
cleaner than in hadronic collisions, and the four momentum of the $B$
is precisely determined. Moreover these decay products are very massive
and only a small amount ($\simeq 128$ MeV) of the initial energy is
converted into kinetic energy. The decay particles move slowly in the
$B$ rest frame, interact strongly and can thus form bound states just
below threshold or resonances just above threshold.

The CLEO collaboration has already provided a measurement of the
inclusive momentum distributions of the $J/\psi$ and $\psi^{'}$ in $B$
decays\cite{cleo}. If the $ \Lambda - \overline p $ is formed as a
bound state, the $ J/\psi $ will be produced with a momentum just above
$0.56$ GeV/c (in the $B$ rest frame). Similarly, if the $ J/\psi -
\Lambda $ is bound, the $\overline p $ will be produced at a momentum
above $0.45$ GeV/c. Finally, if the $ J/\psi - \overline p $ is formed,
the $\Lambda $ will have momentum slightly larger than $0.48$ GeV/c. If
the $B_{\psi}$ state is below the $ J/\psi - p $ mass threshold it may
be observed in the $B^+$ decay as a narrow peak in the $\overline
\Lambda$ momentum distribution at $0.48$ GeV/c. Another interesting
decay channel is shown in Fig. 1b, where the proton and the $D^+_s$
meson may be bound forming the $u\,u\,d\,c\,\overline s$ pentaquark.
The conjugate process is shown in Fig. 1c. In Fig. 1d we illustrate the
$B^-\rightarrow D^-_s\, p \,\overline p $ decay. Here the proton and
the $D^-_s$ meson may be bound forming the $u\,u\,d\,\overline c \, s$
pentaquark. This state, although CKM suppressed, is interesting because
it is more strongly bound than $u\,u\,d\,c\,\overline s$
\cite{moin,lip}. It thus seems possible, with high enough statistics,
to identify two-body structures in these spectra, direct signatures for
the formation of exotic QCD bound states.

%

In Fig. 2 we reproduce the $ B\rightarrow J/\psi X $ spectrum presented
in \cite{cleo}, in which the feed-down modes $B \rightarrow \psi^{'} X$
and $B\rightarrow \chi_{c1}X$ have been subtracted. This procedure
eliminates the $J/\psi$'s coming from decays of excited charmonium
states. The number of remaining points is not large, but we can
nevertheless clearly see a distinct secondary bump in the lower
momentum region corresponding to the $B\rightarrow J/\psi\, \overline
p\, \Lambda$ decay.

A baryonium state such as $ \Lambda - \overline p $ has the same
quantum numbers as an excited kaon. We can distinguish a nuclear-type $
\Lambda - \overline p $ bound state by the fact that its mass should be
within 10's of MeV of the threshold.  Thus $B$ decays offer not only
the possibility of observing baryonium resonances
\cite{ven,bau,arm,cle,apo}, but also bound-state baryonia such as the $
\Lambda - \overline p $ with mass below the threshold ( $2.05$ GeV).
The ($ \Lambda - \overline p $) state recoils against the $J/\psi$ with
momentum (in the $B^-$ c.m.s.) slightly beyond the upper limit ($0.56$
GeV/c) of the phase space for the three body decay $B\rightarrow
J/\psi\, \Lambda \,\overline p $.

If there is a narrow peak at $p_{J/\psi} \simeq 0.56$ GeV/c, the mass
$M_X$ of the $J/\psi$ decay partner can be immediately inferred from
the simple two body phase space formula:
\begin{eqnarray}
p_{J/\psi} = \frac{ M_B}{2} \lambda^{1/2}\left(1,\frac{M^2_X}{M^2_B},
\frac{m^2_{J/\psi}}{M^2_B} \right)
\label{eq:two}
\end{eqnarray}
with
\begin{eqnarray}
\lambda \left(x,y,z\right) & = & x^2 +y^2 +z^2 -2xy -2xz -2yz
\end{eqnarray}
In the above expression $M_B$ is the B meson mass ($= 5.279$ GeV) and
$m_{J/\psi} = 3.097$ GeV. This simple equation can be solved for $M_X$
giving $M_X \simeq 2.05 $ GeV, which is exactly equal to the sum of the
$\Lambda$ and $\overline p$ masses. A peak above $0.56$ GeV would be an
indication of a $\Lambda\,\overline p$ bound-state formation. If the
binding energy of the $ \Lambda - \overline p $ system is $\Delta
m_{\Lambda\,\overline p}\,\simeq 10$ MeV then
\begin{eqnarray}
\Delta p^2_{J/\psi} & = & \Delta m^2_{\Lambda\,\overline p} \cdot
\frac{m_{J/\psi}}{M_B}
\label{eq:resol}
\end{eqnarray}
giving $\Delta p_{J/\psi} = 21 $ MeV/c beyond the upper limit of the
allowed kinematical region for $B\rightarrow J/\psi\, \Lambda
\,\overline p$. Given that the estimated experimental resolution is of
$\Delta p = 1.7$ MeV \cite{bab}, it seems possible to detect such peak.

In Ref. \cite{cleo} the $\psi^{'}$ spectrum is also presented. The
statistics are still small, and there is no evidence of any structure.
This is in agreement with our phase-space based considerations:
because of its larger mass ($\simeq 3.686$ GeV), the $\psi^{'}$ decay
partners can have masses of at most $1.593$ GeV. This is not enough to
produce two baryons and therefore a near-threshold decay is excluded.

\section{Modified phase-space analysis of $B$-decay}

Quantitative predictions for exclusive decay amplitudes such as $B
\rightarrow J/\psi \Lambda \bar p$ or strange baryonium formations $B
\rightarrow J/\psi [\Lambda\bar p]$ involve all of the complexities and
uncertainties of QCD hadronization at the amplitude level and at low
relative momentum where the gauge interactions are strongest.
Nonrelativistic QCD (NRQCD) has brought significant progress to this
field. The underlying weak decay $b \rightarrow c\bar cs$ provides a
guide to the flow of the heavy quarks. Very recently \cite{pal} the
$J/\psi$ momentum distribution in the decay $B\rightarrow J/\psi\,+\,X$
was calculated for the first time and compared to data. It is
interesting to observe that the obtained spectra agree with data, but
fail in the low momentum region, exactly where we may expect some
enhancement due to the exclusive channel considered here.

In general, hadronization will distort the ``bare'' quark momentum
distribution given by the weak matrix element. Another source of
uncertainty is the momentum distribution of the initial quarks inside
the $B$ meson, which was shown in \cite{pal} to strongly affect the
spectrum of final particles. In order to keep our approach simple, we
shall assume that the $J/\psi$ spectrum measured over the whole
available phase space gives a good indication of how the complicated
underlying dynamics modifies the pure phase space picture ( $|{\cal
M}|^2 = $ const) , thus providing a good representation of the squared
matrix element.

In what follows we describe our modified phase space analysis of $B$
decay. Following the standard description of the kinematics\cite{kaj}
the decay rate, at the hadronic level, is given by:
\begin{eqnarray}
d \Gamma & = & \frac{1}{2\sqrt{s}} \frac{1}{2}\sum |{\cal M}|^2 \left(2
\pi \right)^{4-9} \prod_{i=1}^{3} \frac{d^3 p_i}{2 E_i} \delta^3
\left(\vec{p}-\vec{p_1}-\vec{p_2}-\vec{p_3}\right) \delta
\left(\sqrt{s}-E_1-E_2-E_3\right).
\label{eq:rate}
\end{eqnarray}
In the above expression the indexes 1,2 and 3 refer to the $J/\psi$,
$\Lambda$ and $\overline p $ respectively. $E_{i}$ is the energy of the
$i^{th}$ particle and $\sqrt{s}$ is the center of mass energy.  We take
$|{\cal M}|^2 $ as
\begin{eqnarray}
|{\cal M}|^2 \longrightarrow {\rm const} \cdot f(p_1)
\label{eq:matf}
\end{eqnarray}
where the function f is given by
\begin{eqnarray}
f(x) & = & x\, \cdot \exp \left( - \frac{ \left( x- x_0 \right)^2}
{\sigma_0^2}\right) \cdot \left( 1- x \right)
\label{eq:f}
\end{eqnarray}
where $x = p_1\, /\, p_{\rm max}$, $x_0 = 2.0\, /\, p_{\rm max}$,
$\sigma_0 = 0.92\, / \,p_{\rm max}$ and $p_{\rm max}$ is the maximum
momentum allowed for the $J/\psi$. In our case $p_{\rm max} = 1.95$ GeV
and the function f reproduces the data presented in \cite{cleo}.  The
normalization of $f$ is absorbed in the constant in eq. (\ref{eq:matf}).

In the rest frame of the decaying particle ($\vec{p}=0$ and $\sqrt{s} =
M_B $) the integration over the variables 2 and 3 can be then easily
done and we obtain the $J/\psi$ momentum spectrum:
\begin{eqnarray}
\frac{d \Gamma}{d p_1} & = & \frac{p_1}{E_1} f(p_1) \frac{1}{s_2}
\lambda^{1/2}\left(M^2_B,s_2,m^2_1\right)
\lambda^{1/2}\left(s_2,m^2_2,m^2_3\right)
\label{eq:spec}
\end{eqnarray}
where we have omitted all constant factors (since the normalization is
free) and $s_2  =  M^2_B +m^2_1 - 2 M_B E_1 .$ Inserting in eq.
(\ref{eq:spec}) the values of the masses, we obtain a spectrum which is
adjusted to data and plotted in Fig. 2 as a solid line. As can be seen,
the position of the maximum given by the simple modified phase space
model coincides with the maximum of the bump. In Fig. 2 we also show
(with a dotted line) the fit of data with eq.(\ref{eq:f}).  In the
region $ p_{J/\psi} \leq 0.56 $~GeV/c this curve is an estimate of the
background, \ie, all processes other than eq. (\ref{eq:bran}) in which
a $J/\psi$ is produced. Subtracting the integral of the dotted curve
(over $ 0 \leq p_{J/\psi} \leq 0.56 $~ GeV/c ) from the integral of the
solid line we can obtain an estimate for the branching fraction for eq.
(\ref{eq:bran}) $\simeq 0.04 \%$.

For a baryonium binding energy of $\Delta m_{\Lambda\,\overline p}\,=
10$ MeV, the exact position of the peak in the $J/\psi$ distribution is
given by eq. (\ref{eq:two}) with $ M_{X}\,=\,m_{\Lambda}+m_{\overline
p}-\Delta m_{\Lambda\,\overline p} $. We assume that the width of this
state is equal to its binding energy. A final condition must be
satisfied by this bound state: it will be observed only if the strength
of the peak is much larger than the background.  For illustration we
have estimated this strength to be $10 \%$ of the whole distribution.
This produces a clear signal, which is shown in Fig. 2 with a dashed
line. More details are given in the next section.

\section{The strength of the peak}

In this section we explain how to estimate the strength of the
quasi-two-body decays corresponding to the area below the (dashed line)
peak in the $J/\psi$ momentum distribution. The process
\begin{eqnarray}
B^-\longrightarrow J/\psi\, \Lambda \,\overline p
\label{eq:hadronic}
\end{eqnarray}
can be understood, at the quark level, as the reaction
\begin{eqnarray}
b\,+\overline u\,\longrightarrow\, c\, \overline c\, u\, d\,
s\, \overline u\, \overline u\, \overline d
\label{eq:partonic}
\end{eqnarray}
characterized by a cross section, $\sigma$, which cannot be rigorously
calculated. Since we are always treating the case where we have a
$J/\psi$ , we can coalesce the two charmed quarks into one particle and
say that the effective reaction is
\begin{eqnarray}
b\,+\overline u\,\longrightarrow\, J/\psi\, +\, u\, d\, s\,
\overline u\, \overline u\, \overline d
\label{eq:partonic1}
\end{eqnarray}
where the six quarks system has invariant mass $m_6$. Following the
spirit of the approximations made in the last section, the differential
cross section $\sigma$ can be written as
\begin{eqnarray} d \sigma &=& {\rm const}\, \cdot \, d R_7
\label{eq:sigr7}
\end{eqnarray}
where the constant factor represents the matrix element squared and
$R_7$ is the phase space factor for the seven final particles ($
J/\psi\,\, u\,\, d\,\, s\,\, \overline u\,\, \overline u\,\, \overline
d $ ), which, with the help of a recursion relation \cite{kaj}, can be
written as a function of $m_6$ :
\begin{eqnarray}
d R_{7} & = & \frac{d^3 p_{J/\psi}}{2 E_{J/\psi}} R_{6} (m_{6})
\label{eq:r7}
\end{eqnarray}
In the center of mass system, from energy momentum conservation we have
\begin{eqnarray} M_B\, &=& \, \sqrt{p^2_{J/\psi} + m^2_{J/\psi} } +
\sqrt{p^2_{J/\psi} + m^2_6}
\label{eq:relm}
\end{eqnarray}
Solving this relation for $p_{J/\psi}$, differentiating with
respect to $m_6$, substituting in eq. (\ref{eq:r7}) and inserting eq.
(\ref{eq:r7}) into eq. (\ref{eq:sigr7}) we obtain
\begin{eqnarray}
\frac {d \sigma}{d m_{6}} & \simeq & \sqrt{M_B - m_{J/\psi} - m_{6} }
\left(\frac{m_{6}} {m_{J/\psi} + m_{6}} \right)^{3/2} R_{6} (m_{6})
\label{eq:r7new}
\end{eqnarray}
where a nonrelativistic approximation is made and some immaterial
constants are dropped. With the additional assumption that the six body
phase space is approximately one-dimensional and taking constituent
masses for the quarks ($\simeq 300$ MeV) we can evaluate $R_{6}$
(almost) analytically.  As expected the curve Eq. (\ref{eq:r7new}) has
a bell shape. $R_{6}$ starts from zero when $m_{6}$ is just the sum of
the quark masses and then increases monotonically. The multiplying
factor in Eq. (\ref{eq:r7new}) is a decreasing function of $m_{6}$ and
goes to zero when it is equal to $M_B - m_{J/\psi}$. This cross section
can be integrated over different $m_6$ intervals. The bound-state
region corresponds to $m_6$ smaller than the $\Lambda-\overline p$
threshold, \ie,
\begin{eqnarray}
m_p + m_{\Lambda} - \varepsilon \leq m_{6} \leq m_p + m_{\Lambda}
\label{eq:bound}
\end{eqnarray}
where $\varepsilon$ is the binding energy. The continuum region is
given by
\begin{eqnarray}
m_p + m_{\Lambda} \leq m_{6} \leq M_B - m_{J/\psi}
\label{eq:cont}
\end{eqnarray}
The region
\begin{eqnarray}
m_{6} \leq m_p + m_{\Lambda} - \varepsilon
\end{eqnarray}
refers to processes in which the $B$ mesons decays into pions and a
kaon with too small invariant mass.  Such events do not come from the
baryonium decay and are thus uninteresting. This part of the phase
space will be ignored.

The probability of forming a bound state is then given by
\begin{eqnarray}
p\,&=&\, \frac {\int_{ m_p + m_{\Lambda} - \varepsilon}^{m_p +
m_{\Lambda}}  d m_{6} \frac{d \sigma}{d m_{6}}} {\int_{ m_p +
m_{\Lambda} - \varepsilon}^{M_B - m_{J/\psi}} d m_{6} \frac {d
\sigma}{d m_{6}}}
\label{eq:prob}
\end{eqnarray}
We use $\varepsilon = 10$ MeV. Performing the calculations we find that
$p\,=\,10 \% $.

On the other hand, looking at Fig. 2, we observe that the probability
of forming a bound state may be identified with the ratio between the
area below the (dashed line) peak and the total area (solid plus dashed
lines), \ie, the differential branching fraction integrated over
$p_{J/\psi}$ :
\begin{eqnarray}
p\,&=&\, \frac{ \int_{p_{\rm max}}^{p_{\rm max}+ \Delta
p_{J/\psi}} \frac{d B}{dp_{J/\psi}} d p_{J/\psi}} {\int_{0}^{p_{\rm
max}+\Delta p_{J/\psi}} \frac{d B}{d p_{J/\psi}} d p_{J/\psi}}
\label{eq:probnew}
\end{eqnarray}
where $\Delta p_{J/\psi}$ given by Eq. (\ref{eq:resol}). Knowing $p$
from eq. (\ref{eq:prob}), we replace it in the above expression and
determine the right hand side, \ie, the strength of the peak, which,
with the parameters stated above, is $10 \% $.

The same procedure can be applied to the other interesting bound states
in the decay mode (\ref{eq:bran}) as mentioned at the end of the
introduction. The ($ J/\psi-\overline p $) and ($ J/\psi-\Lambda $) may
be observed as peaks in the momentum distribution of the $\Lambda$ and
$\overline p $ respectively. These spectra are computed with the help
of eq. (\ref{eq:spec}). Now, $p_1$ refers to $\overline p $ or to
$\Lambda$ and $f(p_1) = 1 $, since we do not know how to modify the
phase space in these cases where no data are available. The strengths
of the bound state peaks are computed in the way described above (with
$\varepsilon = 10$ MeV and $\Delta m_{J/\psi\,\overline p}\,=\,\Delta
m_{J/\psi\,\Lambda}\,=\,  10$ MeV), and the results are $ p = 10.7 \% $
and $ p = 11 \% $ for the $\Lambda$ and $\overline p $ spectra
respectively. These spectra (with corresponding peaks) are shown in
Fig. 3 in arbitrary units.

\section{Conclusions}

The high luminosity of the CESR storage ring and the $B$-factories now
under construction will provide a high statistic sample of $B$ decays.
As we have discussed in this paper, the decay of the $B$ may lead to
the production of exotic QCD states such as quarkonium-baryonium bound
states, pentaquarks, and strange baryonium.  The nearly monoenergetic
production of the recoil hadron in the rest system of the $B$ could
provide a sensitive and clean signal for these states.
\bigskip

\underline{Acknowledgments}: F.S.N. would like to thank the SLAC theory
group for the warm hospitality extended to him during his stay there.
It is a pleasure to thank T. Kodama, Y. Hama, G.P. Lepage, H. Quinn,
and G. Miller for helpful discussions. 

\newpage

\noindent {\bf Figure Captions}\\

\begin{itemize}
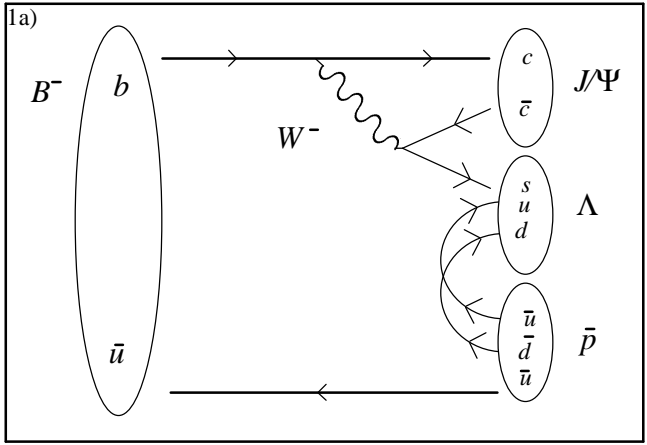
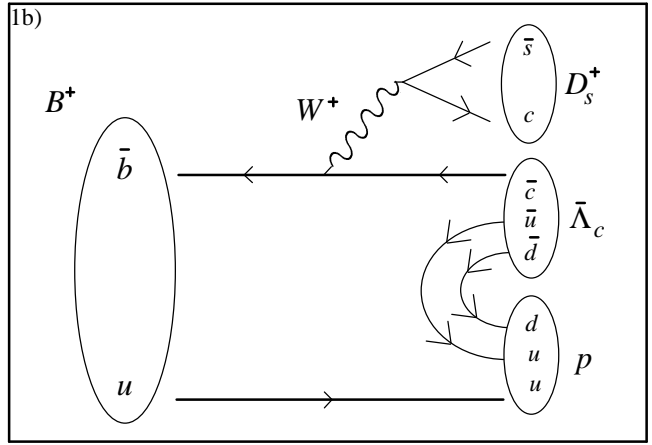
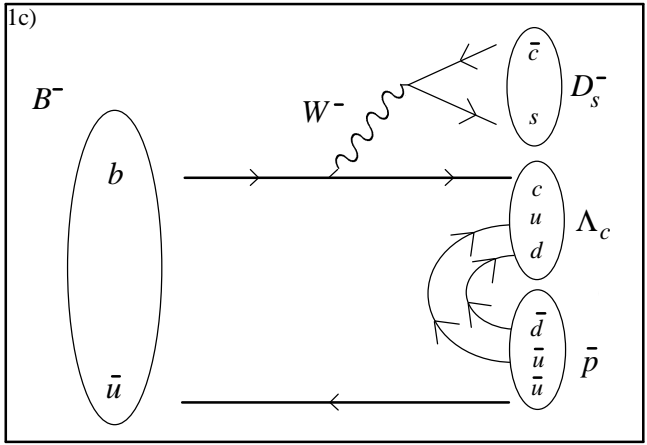
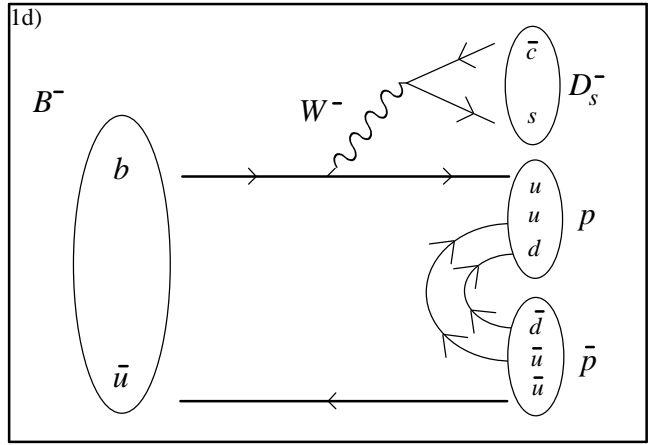
\item[{\bf Fig. 1}]
Exclusive hadronic $B$ decays which may lead to the formation of bound
states. 1a) hadronically bound $J/\psi$ or strange baryonium; 1b)
$u\,u\,d\,c\,\overline s$ pentaquark; 1c) $\overline u\, \overline
u\,\overline d\,\overline c\, s$ pentaquark; 1d) $u\,u\,d\,\overline
c\, s$ pentaquark.

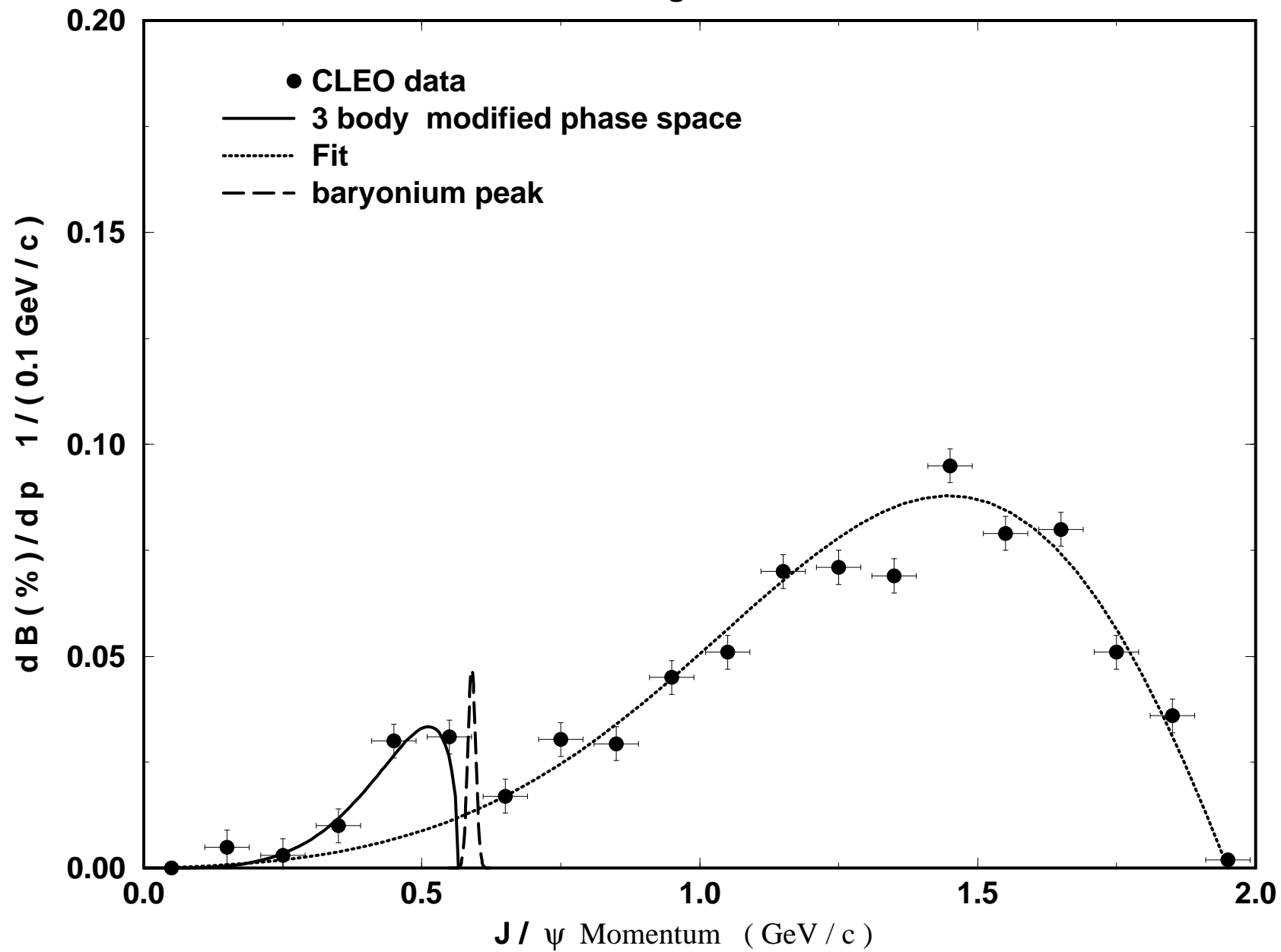
\item[{\bf Fig. 2}]
The $J/\psi$ momentum spectrum measured by the CLEO Collaboration
\cite{cleo} after the subtraction of the feed-down modes. The solid
line is the result of a modified phase space calculation for the
$B\rightarrow J\psi\,\Lambda\,\overline p$ three-body decay,
eq.(\ref{eq:spec}). The dotted line shows the fit of data,
eq.(\ref{eq:f}), used to estimate background contributions. The dashed
line represents the expected peak if there is a narrow strange
baryonium with binding energy of $10$ MeV and width $10$ MeV.

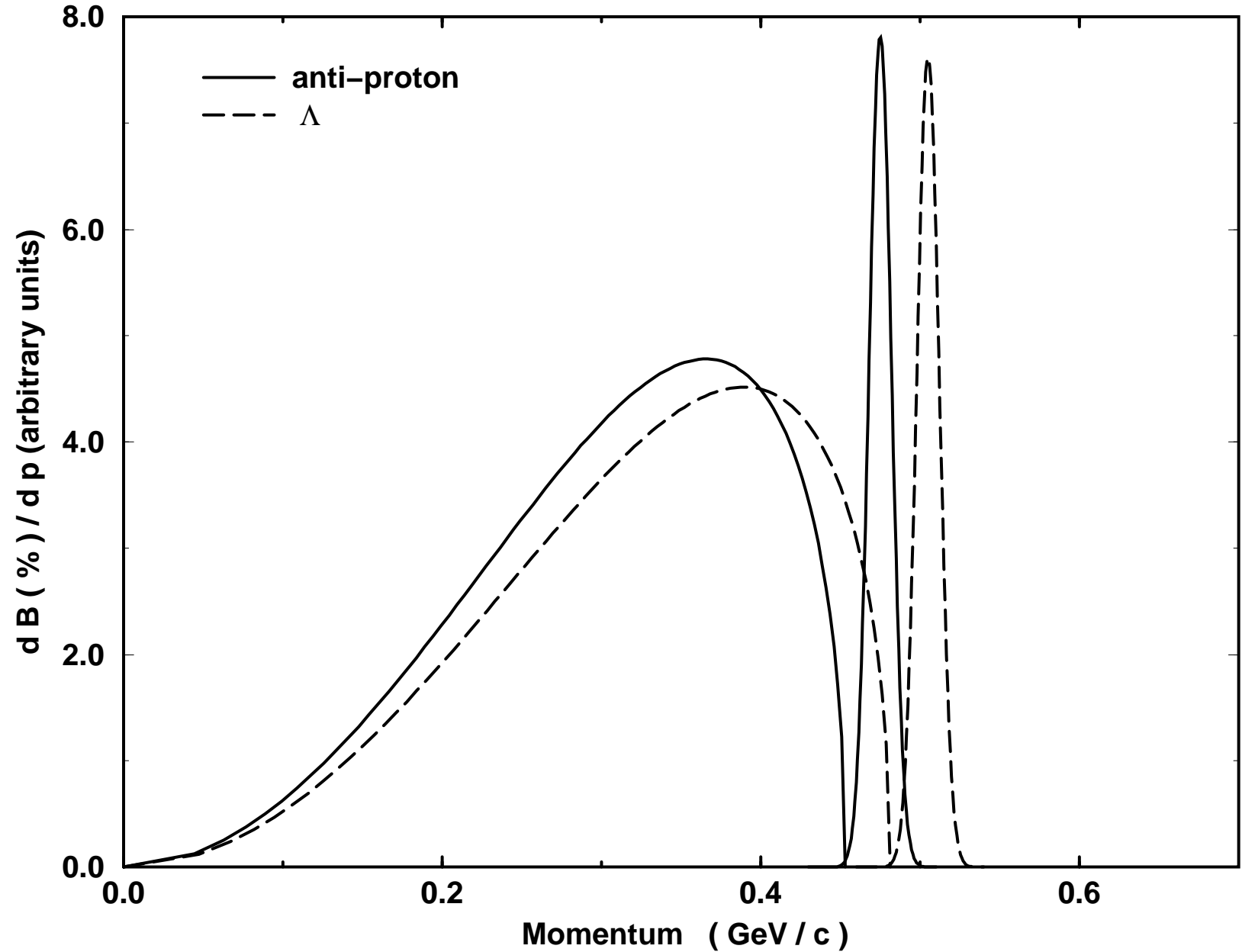
\item[{\bf Fig. 3}]
The momentum distribution of the $\overline p$ (solid line) and
$\Lambda$ (dashed line) produced in the $B$ decay eq.(\ref{eq:bran}).
The broad curves show the three-body decay spectra of
the $\overline p$ and $\Lambda$ calculated with eq.(\ref{eq:spec}).
Just beyond the kinematical limits of these two curves we show the
peaks which would be signatures of the below-threshold
($ J/\psi-\Lambda $) (solid line)
and ($ J/\psi-\overline p $) (dashed line) bound states.
Their binding energies and widths are assumed to be
$\varepsilon = 10 $ MeV. These numbers determine the position
and width of the bound state peaks. Their strengths are given
by eqs. (\ref{eq:prob}) and (\ref{eq:probnew}).

\end{itemize}

\end{document}